\documentclass[a4paper]{jpconf}
\usepackage[dvipdfmx]{graphicx}
\usepackage{afterpage}
\begin{document}

\def\ABT{$\alpha$-(BEDT-TTF)$_2$I$_3$}
\def\ABTs{$\alpha$-(BEDT-TTF)$_2$I$_3$\ }

\title{Anisotropic spin motive force in multi-layered Dirac fermion
system, \ABT}

\author{K Kubo and T Morinari}

\address{Graduate School of Human and Environmental Studies, Kyoto
University, Kyoto 606-8501, Japan}

\ead{kubo.kenji.22x@st.kyoto-u.ac.jp}

\begin{abstract}
We investigate the anisotropic spin motive force in \ABT, which is a
 multi-layered massless Dirac fermion system under pressure. Assuming the
 interlayer antiferromagnetic interaction and the interlayer anisotropic
 ferromagnetic interaction, we numerically examine the spin ordered
 state of the ground state using the steepest descent method. The
 anisotropic interaction leads to the anisotropic spin ordered state. We
 calculate the spin motive force produced by the anisotropic spin
 texture. The result quantitatively agrees with the experiment.

\end{abstract}

\section{Introduction}
An organic conductor \ABTs is a multi-layered massless Dirac fermion
system, in which conduction layers of BEDT-TTF molecules and I$_3$
anions stack alternatively. In each conduction layers, the massless
Dirac fermion system is realized. The valence band and the conduction
band contact at two inequivalent points in the
Brillouin zone. The energy dispersion is linear in the vicinity of the
contact points. The Fermi energy coincides with the energy of the contact point. Therefore, the system is called Dirac fermion system.

Under the magnetic field $B$, the energy of the $n$-th
Landau level of the Dirac fermion is described by $E={\rm
sgn}(n)C\sqrt{|n|B}$ with a constant $C$. The energy difference between the first
and the zeroth Landau levels is $C\sqrt{B}$. Therefore, in relatively high
magnetic field, we may consider only the Landau level with $n=0$. 

\ABTs has unique spin ordered states [1]. In each layer, the quantum Hall
ferromagnetic state is realized by the intralayer ferromagnetic
interaction due to the exchange interaction. On the other hand, the
interlayer ferrimagnetic state is possible by the interlayer antiferromagnetic interaction.

\begin{figure}[h]
 \begin{center}

  \begin{tabular}{cc}
    \begin{minipage}{14pc}
     \begin{center}
      \includegraphics[width=12pc]{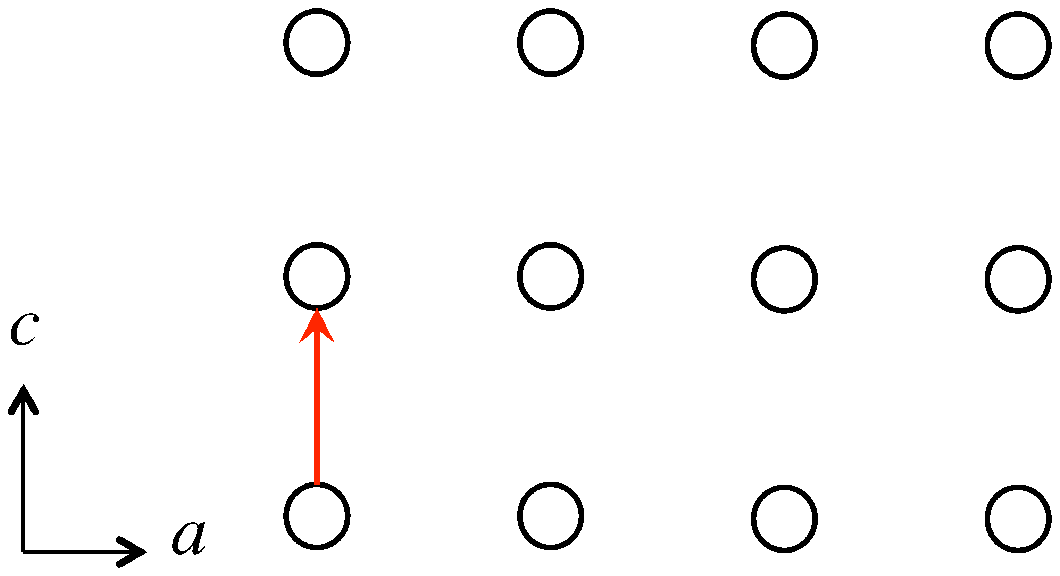}\\
     ($a$)
     \end{center}
    \end{minipage}
   \hspace{2pc}&
   \begin{minipage}{14pc}
    \begin{center}
     \includegraphics[width=14pc]{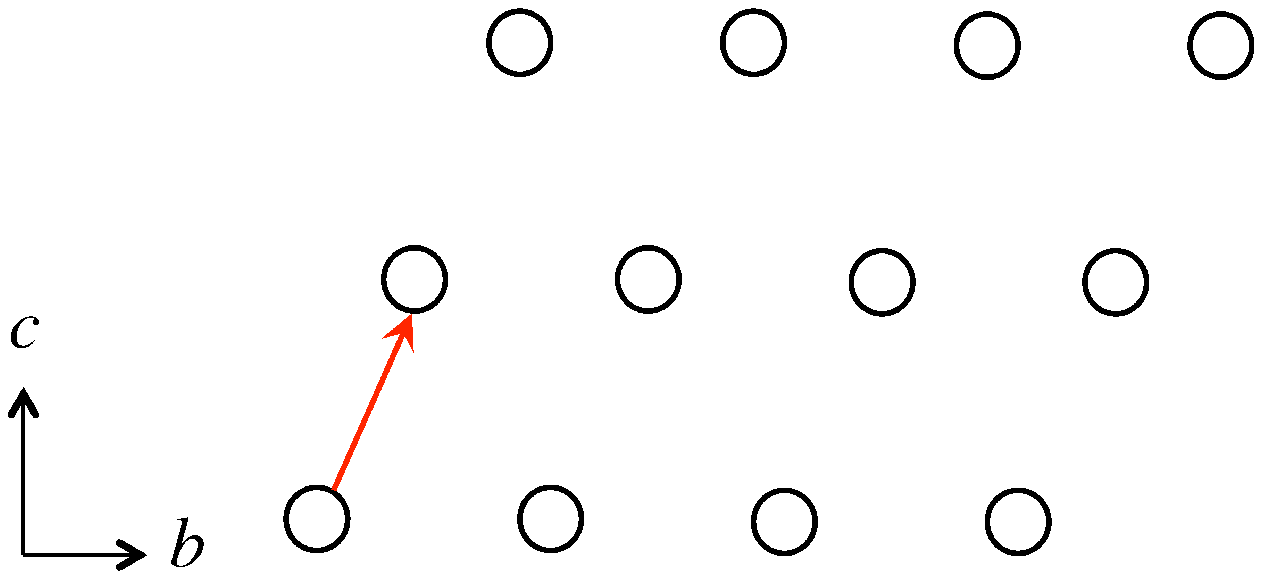}\\
    ($b$)
    \end{center}
    \end{minipage}

  \end{tabular}
 \end{center}
 \caption{\label{direction}The red arrow shows the interlayer tunneling direction. The
 BEDT-TTF molecules are described by the circles. $a$- and $b$-axis is in the
 conducting plane. $c$-axis is perpendicular to it. ($a$) The interlayer tunneling direction in $a$-$c$ plane. ($b$) The interlayer tunneling in $b$-$c$ plane is tilted in the direction to the $b$-axis.}
\end{figure}

\ABTs has anisotropic crystal structure and transport property. In
figure \ref{direction}, the direction of the interlayer tunneling is
shown. The tunneling is tilted in the direction of $b$-axis. For the
experimental data, the angle between the vertical direction and the
interlayer tunneling direction is about $28^\circ$.

At low temperature and under magnetic field, anisotropic voltage is
observed in \ABTs[2]. At high temperature the voltage disappears. The
voltage increases with increasing the magnetic field. Therefore, one
possible scenario is that the voltage is caused by spin motive force. The anisotropic
crystal structure can lead to in the anisotropic voltage. 
The spin motive force produces the voltage. However, this is not a
equilibrium situation. The spin motive force is transient due to the
Gilbert damping. Charge accumulates at the edge of the system by the
spin motive force and produces a electric field opposite to that created
by spin motive force. The voltage will vanish due to the electric field
created by the charge accumulation. Hereafter, for simplicity, we assume
that the necessary time of the disappearance of the voltage is long and the voltage is constant
in this time scale.

In this paper, we show that anisotropic spin ordered states are
realized in \ABTs because of the interlayer anisotropic interaction and
anisotropic spin motive forces are produced under magnetic field.

\begin{figure}[h]
 \begin{center}
  \includegraphics[width=14pc]{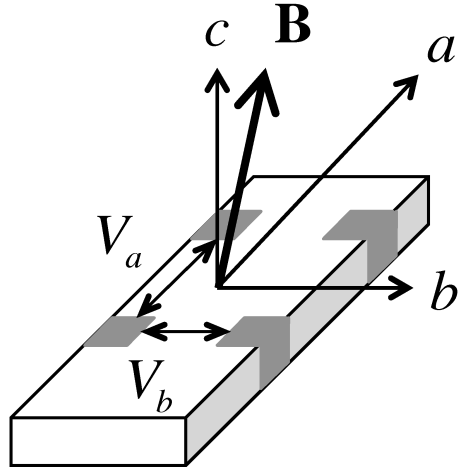}
 \end{center}

 \caption{\label{voltage}The anisotropic voltage in \ABT. The voltage in the direction
 of $a,b$ are defined by $V_a,V_b$.}
\end{figure}

\section{Model}

We assume the square lattice for the conducting plane and the ferromagnetic
interaction $J_{\rm intra}$ is assumed between nearest neighbor molecules. We define $b$- and
$a$-axis as $x$- and $y$-axis, respectively. We take the direction of interlayer
tunneling as $z$-axis. The interlayer interaction in the direction of 
tunneling is denoted by $J_{\rm inter}<0$. In this system, the lattice is distorted in $x$-$z$ plane since the molecular arrangement is
tilted in the direction of $x$-axis and there are two different next
nearest neighbor sites in $x$-$z$ plane. When the distance is different,
the interaction should be different. For this reason, we assume a
finite interaction $J'_{\rm inter}$ in one of the next nearest neighbor
sites in $x$-$z$ plane. On the other hand, for simplicity, we take a
square lattice in $x$-$z$ plane in this calculation and assume the magnetic field is parallel to the $z$-axis. 

\begin{figure}[h]
 \begin{center}
  \begin{tabular}{cc}
   
   \begin{minipage}{14pc}
    \begin{center}
     \includegraphics[width=14pc]{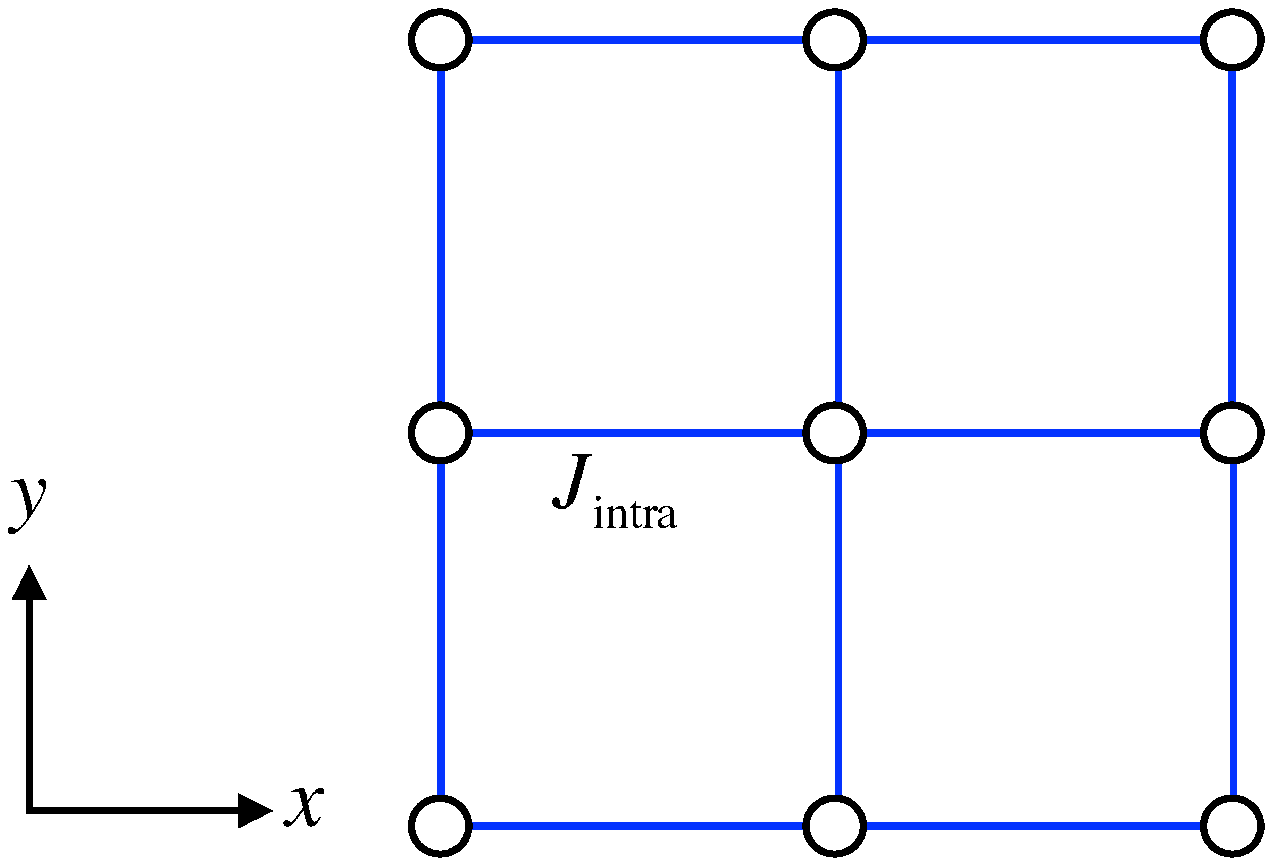}\vspace{1pc}\\
     ($a$)
    \end{center}
   \end{minipage}
   \hspace{2pc}&
   \begin{minipage}{14pc}
    \begin{center}
     \includegraphics[width=14pc]{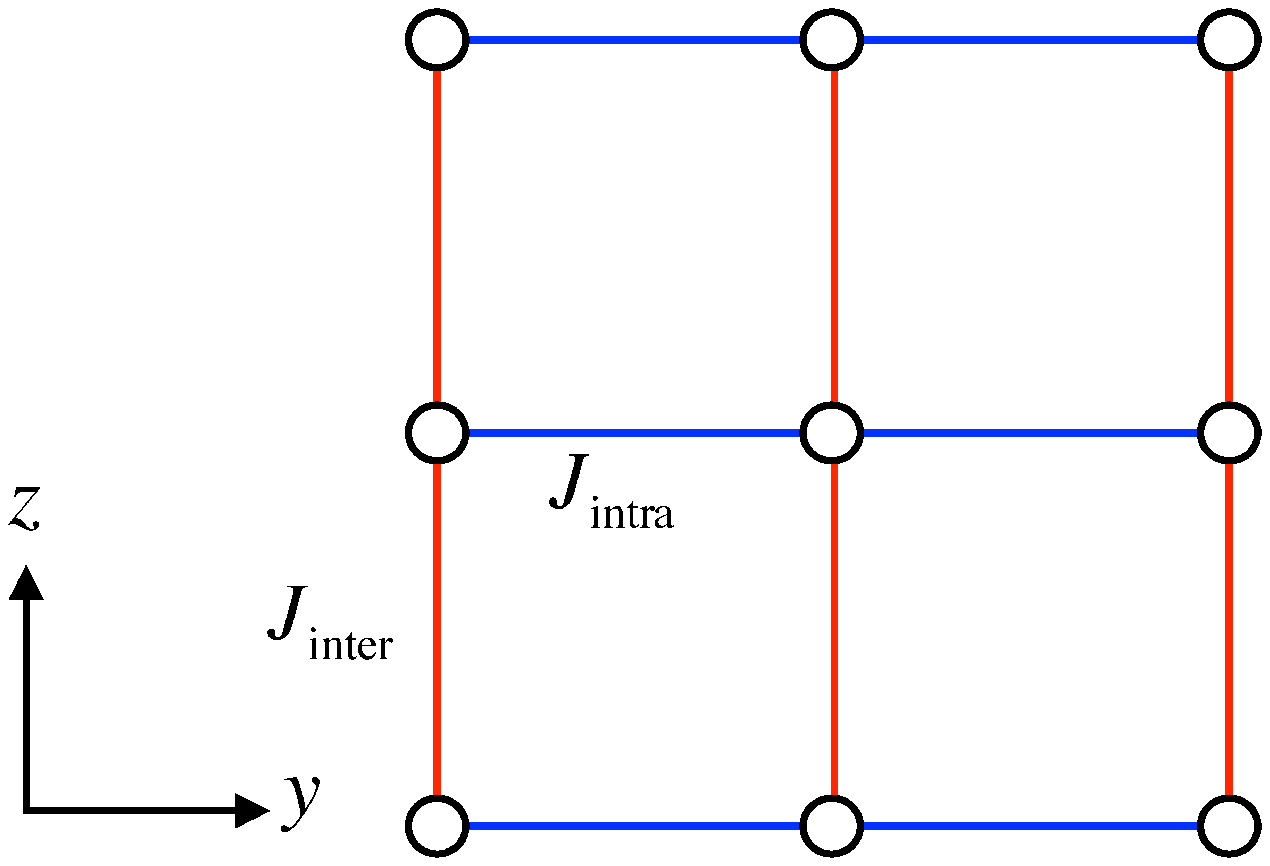}\vspace{1pc}\\
     ($b$)
    \end{center}
   \end{minipage}
   \vspace{2pc}
   \\

   \begin{minipage}{14pc}
    \begin{center}
     \includegraphics[width=14pc]{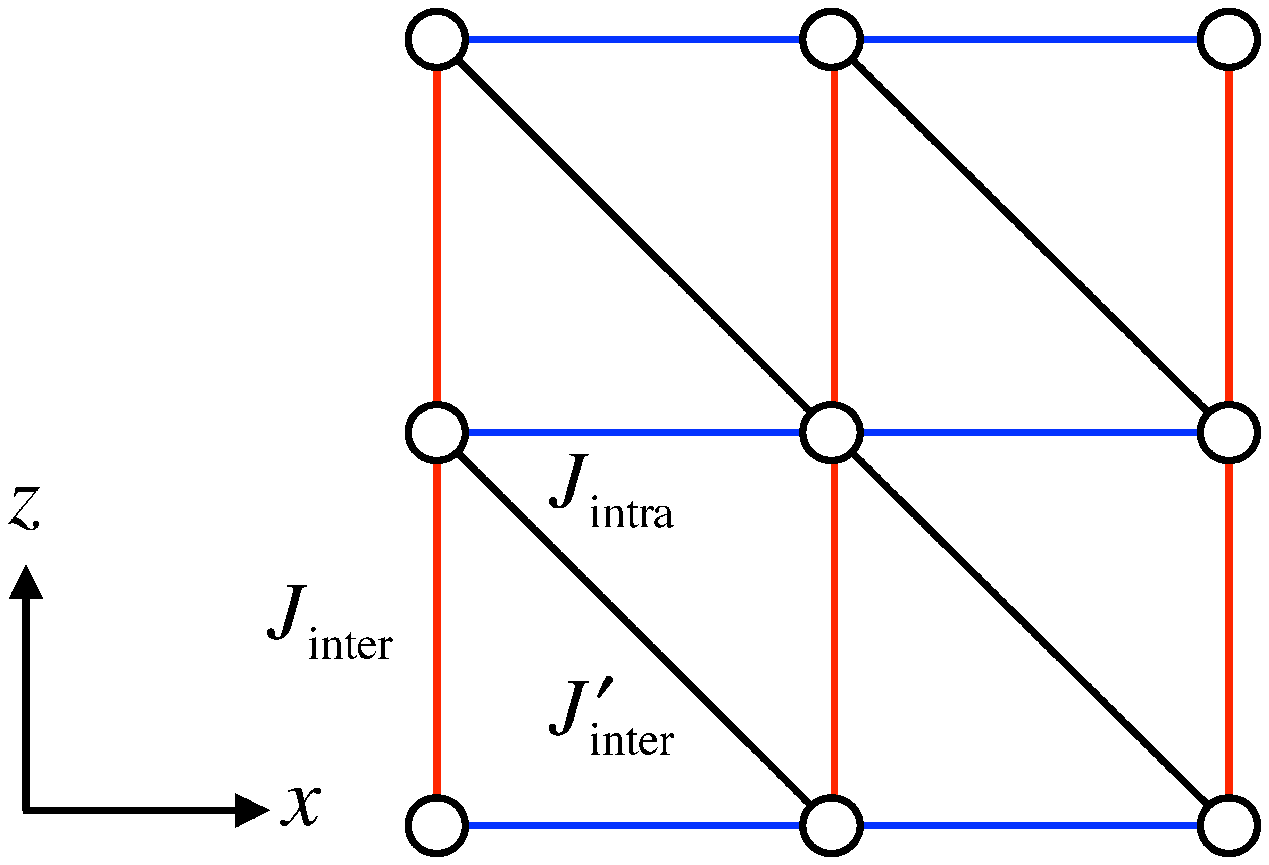}\vspace{1pc}\\
     ($c$)
    \end{center}
   \end{minipage}
   \hspace{2pc}&
   \begin{minipage}{14pc}
    \caption{The model of \ABT. ($a$) Only $J_{\rm intra}$ exists in $x$-$y$
 plane. ($b$) $J_{\rm inter}$ is along $z$-axis and $J_{\rm intra}$ is along
 $y$-axis in $y$-$z$ plane. ($c$) $J_{\rm inter}$ is along $z$-axis,
 $J_{\rm intra}$ is along $x$-axis and $J'_{\rm inter}$ is between one
 side of the next nearest neighbor sites in $x$-$z$ plane.}
   \end{minipage}
  \end{tabular}
 \end{center}
\end{figure}

The Hamiltonian under the magnetic field is written as
\begin{eqnarray}
 H&=&J_{{\rm intra}}\sum_{i,j,k}\left(\mathbf{S}_{i+1,j,k}+\mathbf{S}_{i,j+1,k}+\mathbf{S}_{i-1,j,k}+\mathbf{S}_{i,j-1,k}\right)\cdot\mathbf{S}_{i,j,k}\nonumber\\&+&J_{{\rm
  inter}}\sum_{i,j,k}\left(\mathbf{S}_{i,j,k+1}+\mathbf{S}_{i,j,k-1}\right)\cdot\mathbf{S}_{i,j,k}+J'_{{\rm
  inter}}\sum_{i,j,k}\left(\mathbf{S}_{i,j+1,k+1}+\mathbf{S}_{i,j-1,k-1}\right)\cdot\mathbf{S}_{i,j,k}\nonumber\\
 &-&\frac{1}{2}g\mu_B\sum_{i,j,k}\mathbf{B}\cdot\mathbf{S}_{i,j,k},
\end{eqnarray}
where $\mathbf{S}_{i,j,k}$ is a spin at the lattice point $(i,j,k)$. $g$
is g factor and $\mu_B$ is the Bohr magneton.

In this model, we investigate the spin ordered state of the ground state
using the steepest descent method. At first, we put a spin with random
direction on each lattice point. Next, we update spins with
\begin{eqnarray}
 \mathbf{n}_{i,j,k}&=&-\frac{J_{{\rm
  intra}}}{4}\left(\mathbf{n}_{i+1,j,k}+\mathbf{n}_{i,j+1,k}+\mathbf{n}_{i-1,j,k}+\mathbf{n}_{i,j-1,k}\right)\nonumber\\
 &-&\frac{J_{{\rm
  inter}}}{2}\left(\mathbf{n}_{i,j,k+1}+\mathbf{n}_{i,j,k-1}\right)-\frac{J'_{{\rm inter}}}{2}\left(\mathbf{n}_{i,j+1,k+1}+\mathbf{n}_{i,j-1,k-1}\right) +\frac{1}{2}g\mu_B\mathbf{B}.\nonumber\\
\end{eqnarray}
Here, $\mathbf{n}_{i,j,k}=2\mathbf{S}_{i,j,k}$ is the normalized localize
moment. The resulting converged state is an approximate for the ground
state. $J_{\rm intra}$ is proportional to the $\sqrt{B}$ [1], but, for
simplicity, we assume $J_{\rm intra}$ is constant. We numerically
calculate with the number of sites, $12\times12\times12$ and under the periodic boundary condition.

\begin{figure}[h]
 \begin{center}
  \begin{tabular}{cc}
   \begin{minipage}{14pc}
    \begin{center}
     \includegraphics[width=14pc]{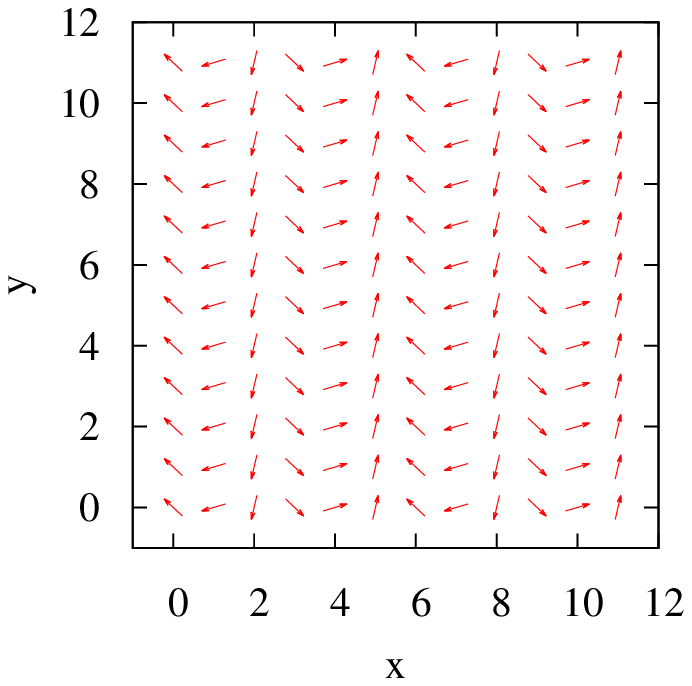}\\
     ($a$)
    \end{center}
   \end{minipage}
   \vspace{2pc}&
   \begin{minipage}{14pc}
    \begin{center}
     \includegraphics[width=14pc]{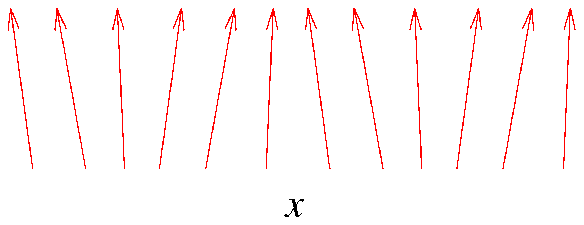}\vspace{1pc}\\
     ($b$)
    \end{center}
   \end{minipage}
  \end{tabular}
  \caption{\label{texture}The resulting spin ordered state of \ABTs with
  $J_{\rm intra}=-5$ K, $J_{\rm inter}=4$ K, $J'_{\rm inter}=-2$ K and $B=5$ T. ($a$) The projection of spins in a layer to $x$-$y$ plane. A periodic spin ordered state is realized along $x$-axis, although a ferromagnetic structure is along $y$-axis. ($b$) The projection of spins in a layer to $x$-$z$ plane.}
 \end{center}
 
\end{figure}

The result with the parameters $J_{\rm intra}=-5$ K, $J_{\rm inter}=4$ K,
$J'_{\rm inter}=-2$ K and $B=5$ T is shown in figure
\ref{texture}. Here, we assume relatively large value for $J_{\rm inter}$, which is the same
order of magnitude as the interlayer hopping estimated in a related
organic compound [3]. A
ferromagnetic state is realized along $y$-axis. On the other hand, a
periodic spin texture is created along $x$-axis due to the interlayer next nearest
neighbor interaction.

Assuming this spin ordered state, we calculate the spin motive force [4] in
the strong coupling limit. In this limit, the spins are polarized and there are no conduction electron with minority spin. Therefore, the spin
motive force is identical with the voltage. We
make indices $j,k$ implicit below since we consider spin motive force in
uniaxial direction. The localized moment is represented as
\begin{eqnarray}
 \begin{array}{ll}
  \mathbf{n}_i=(\cos\theta_i\cos\phi_i,\cos\theta_i\sin\phi_i,\sin\theta_i),
 \end{array}
\end{eqnarray}
where $\theta_i$ is the angle between the direction of the spin and
the $z$-axis. $\phi_i$ is the angle between the direction of the spin
projected in the $x$-$y$ plane and the positive direction of the $x$-axis.
The electric field due to the localized moment is written as
\begin{eqnarray}
 \mathbf{E}_i=\frac{\hbar}{2e}\sin\theta_i\left[\left(\partial_t\theta_i\right)\left(\mathbf{\nabla}\phi_i\right)-\left(\partial_t\phi_i\right)\left(\mathbf{\nabla}\theta_i\right)\right],
\end{eqnarray}
where $e$ is the charge of the electron [5]. Here, the condition
$\partial_t\theta_i=0$ and $\partial_t\phi_i=g\mu_BB/\hbar$ are always
satisfied since the magnetic field is along $z$-axis. The electric field
is rewritten as 
\begin{eqnarray}
 \mathbf{E}_i=-\frac{g\mu_B}{2e}B\sin\theta_i\left(\mathbf{\nabla}\theta_i\right).
\end{eqnarray}
In the numerical calculation with the periodic boundary condition, the
electric field is canceled if we take the average over the spatial period and so the spin motive
force does not appear. However, in the real system, there is the edge of
the sample and a finite spin motive force is possible. On the other hand,
spin motive force along $y$-axis does not appear at all. Therefore, the
spin motive force in this model has strong anisotropy. In order to estimate the spin motive force in the real
system, we calculate the spin motive force with taking a part of the
periodic spin structure. 

\begin{figure}[h]
   \begin{center}
    \includegraphics[width=24pc]{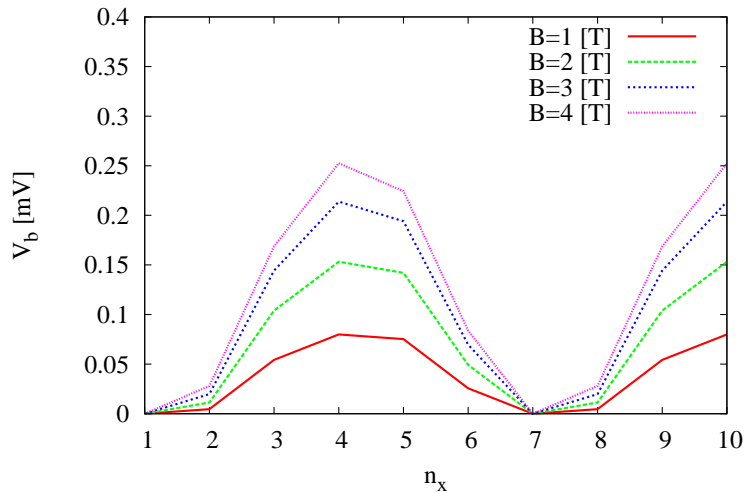}
   \end{center}
 \caption{\label{Vb_B_dependence} Spin motive forces along $x$-axis in
 \ABT. $n_x$ is the number of spins along $x$-axis from the edge of the
 system.}
\end{figure}

\begin{figure}[h]
   \begin{center}
    \includegraphics[width=24pc]{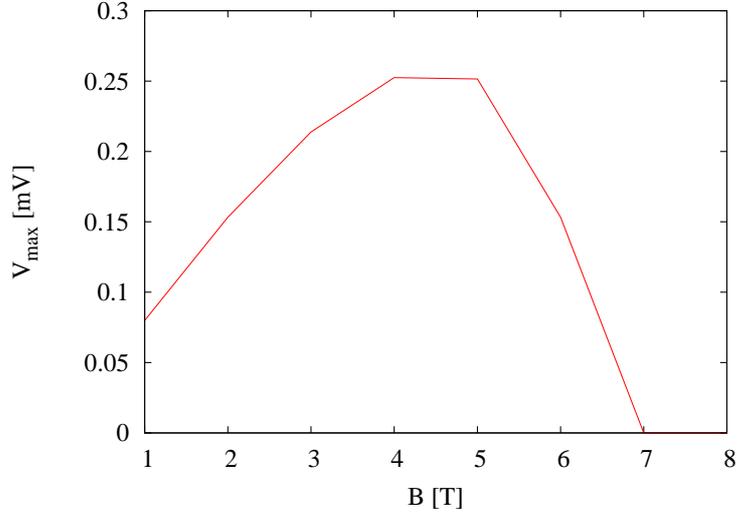}
   \end{center}
 \caption{\label{Vmax} The magnetic field dependence of the maximum
   values of $V_b$. $V_{\rm max}$ increases with increasing magnetic field below
   4 T and decreases above 5 T.}
\end{figure}

The numerical result is shown in figure \ref{Vb_B_dependence}. The
spin motive force along $x$-axis is on the order of $0.01$ mV to $0.1$ mV which is
about the same order compared with the experimental result. The most of
electric field is canceled and only the spin motive force in the
vicinity of the edge remains. The experimental result shows that $V_b$
increases with increasing the magnetic field below $7$ T. In order to
investigate the magnetic field dependence of $V_b$, we define the
maximum value of the $V_b$ as $V_{\rm max}$. Although $V_{\rm max}$ does not
always correspond to the experimental observed voltage, it is able to 
qualitatively evaluate the magnetic field dependence of $V_b$. The
nurerical result of the magnetic field dependence of $V_{\rm
max}$ is shown in figure \ref{Vmax}. $V_{\rm max}$ increases with
increasing magnetic field below $4$ T. However, above $5$ T, $V_{\rm
max}$ decreases. Under high magnetic field, the Zeeman energy is larger
than the intralayer and the interlayer interactions, and so $\{\theta_i\},\{\nabla\theta_i\}$ are small since we assume the
interactions are constant. Therefore, the electric fields due to the
localized moments vanish under high magnetic field. If we consider
the magnetic field dependence of the interactions, we would evaluate the anisotropic spin
motive force more precisely. 

In a real system, the edge shape is different for a different layer. So,
the naive expectation is that if we take the average over those edges,
the spin motive force is canceled out. However, since the anisotropic spin motive force is generated
by the three-dimensional structure, spins in regions without interlayer
interaction should not contribute to the spin motive force. Therefore,
it is justified to consider spin motive force for layers with the same
edge shape.

\section{conclusion}

We have calculated the spin motive force in \ABT. The anisotropic interlayer
interaction exists reflecting the anisotropic crystal structure. The
anisotropic interaction leads to the unidirectional periodic spin
structure. The anisotropic spin motive force is created from the
anisotropic spin structure under magnetic field.

\ack
We would like to thank N. Tajima for discussions and sending us the
experimental result.
This work was financially supported in part by a Grant-in-Aid for
Scientific Research (A) on “Dirac Electrons in Solids” (No. 24244053)
and a Grant-in-Aid for Scientific Research (B) (No. 25287089) and (C)
(No. 24540370) from the Ministry of Education, Culture, Sports, Science
and Technology, Japan.

\end{document}